\documentclass[11pt]{article}
\usepackage{graphicx}
\usepackage[utf8]{inputenc}

\usepackage{amsmath}
\usepackage{amssymb}
\usepackage{epsfig}
\usepackage{graphicx}
\usepackage[numbers,sort&compress]{natbib}
\usepackage{color}
\usepackage[colorlinks=true
,urlcolor=blue
,anchorcolor=blue
,citecolor=blue
,filecolor=blue
,linkcolor=blue
,menucolor=blue
,pagecolor=blue
,linktocpage=true
,pdfproducer=medialab
]{hyperref}

\newcommand\pubnumber{PSI-PR-17-09}
\newcommand\pubdate{\today}

\def\napoli{Paul Scherrer Institut,\\
CH-5232 Villigen PSI, Switzerland}
\def\support{\footnote{This research was supported by the
SNSF under contract 200021\_160156.}}

\def\Title#1{\begin{center} {\Large #1 } \end{center}}
\def\Author#1{\begin{center}{ \sc #1} \end{center}}
\def\Address#1{\begin{center}{ \it #1} \end{center}}

\newcommand\pubblock{\rightline{\begin{tabular}{l} \pubnumber\\
         \pubdate  \end{tabular}}}
\newenvironment{Abstract}{\begin{quotation}  }{\end{quotation}}
\newenvironment{Presented}{\begin{quotation} \begin{center} 
             PRESENTED AT\end{center}\bigskip 
      \begin{center}\begin{large}}{\end{large}\end{center} \end{quotation}}
\def\Acknowledgements{\bigskip  \bigskip \begin{center} \begin{large}
             \bf ACKNOWLEDGEMENTS \end{large}\end{center}}
\def\beq{\begin{equation}}
\def\eeq#1{\label{#1}\end{equation}}
\def\eeqn{\end{equation}}

\def\beqa{\begin{eqnarray}}
\def\eeqa#1{\label{#1}\end{eqnarray}}
\def\eeqan{\end{eqnarray}}

\let\bar=\overbar

\def\Dslash{\not{\hbox{\kern-4pt $D$}}}
\def\dslash{\not{\hbox{\kern-2pt $\del$}}}

\def\msb{{\bar{\ssstyle M \kern -1pt S}}}

\begin{document}
\begin{titlepage}
\pubblock

\vfill
\Title{Charged lepton flavour violation: precise background \\
       calculation and effective field theoretical interpretation}
\vfill
\Author{Giovanni Marco Pruna\support}
\Address{\napoli}
\vfill
\begin{Abstract}
\noindent
This note reviews recent theoretical developments in the study of charged lepton flavour violation. The first part illustrates the status of precise next-to-leading order quantum electrodynamics calculations for the background of charged lepton flavour-violating processes, with a focus on the muonic ``rare'' and ``radiative'' decays. Phenomenological implications of these computations and their impact on present and future experiments will be discussed. The second part describes the recent progress in the effective field theory interpretation of charged lepton-flavour violating observables in connection with different energy scales. A systematic approach is briefly presented and applications on muonic and tauonic observables are reported. This note is submitted as part of the conference proceedings for "NuPhys2016: Prospects in Neutrino Physics".
\end{Abstract}
\vfill
\begin{Presented}
NuPhys2016, Prospects in Neutrino Physics\\
Barbican Centre, London, UK,  December 12--14, 2016
\end{Presented}
\vfill
\end{titlepage}
\def\thefootnote{\fnsymbol{footnote}}
\setcounter{footnote}{0}

\section{Introduction}

Lepton flavour-violating transitions in the charged lepton sector are accidentally forbidden in the Standard Model (SM). Even with the inclusion of neutrino masses, such processes are extremely suppressed and, consequently, any potential signal of charged lepton flavour violation (cLFV) should be interpreted as an indication of new physics (NP). Therefore, an extended experimental programme has been ongoing for decades in the search for fundamental cLFV interactions~\cite{TheMEG:2016wtm,Bellgardt:1987du,Bolton:1988af,Bertl:2006up,Aubert:2009ag,Hayasaka:2007vc,Lees:2010ez,Aaij:2014azz,Aad:2014bca,Akers:1995gz,Abreu:1996mj,Khachatryan:2015kon}.

Recently, systematic efforts have been devoted to understanding the theoretical aspects related to the interpretation of the absence of cLFV signals in terms of limits on beyond the SM (BSM) physics. On the one hand, these studies progressed onto the determination of the next-to-leading order (NLO) quantum electrodynamics (QED) background of cLFV decays, on the other hand they brought a better understanding of the effective field theoretical interpretation of the absence of signals in terms of limits on Wilson coefficients of the SM extended with dimension-six operators (SMEFT).

Concerning the first research line, a precise background calculation was carried out for the so-called ``radiative''~\cite{Fael:2015gua,Fael:2016hnz,Pruna:2017upz} and ``rare''~\cite{Fael:2016yle,Pruna:2016spf} decays, \emph{i.e.} for the processes $l\to l'\gamma+2\nu$ and $l\to 3l'+2\nu$ (or $l\to l'+2l''+2\nu$), respectively.
Both these channels are important\footnote{However, they only represent the fundamental background. The leading source of background consists of accidental fake signals.} for the determination of the limits on the branching ratios (BRs) of the two cLFV processes $l\to l'\gamma$ and $l\to 3l'$ (or $l\to l'+2l''$) because they provide an identical signal in the circumstance where invisible energy tends to zero, especially in view of the new experimental plans to improve the exploring power on these channels by MEG~II~\cite{Baldini:2013ke} and Mu3e~\cite{Blondel:2013ia} for the muonic case, and %BaBar~\cite{Lees:2015gea} and
Belle~II~\cite{Abe:2010gxa} for the tauonic case.

For the phenomenological interpretation of the absence of a signal in terms of limits on the parameter space of potential BSM scenarios, a systematic effective field theory (EFT) treatment was proposed~\cite{Pruna:2014asa,Pruna:2015jhf} and further developed with a particular focus on the muonic coherent LFV transitions~\cite{Crivellin:2016ebg,Crivellin:2017rmk}. This approach consists of expanding the SM (either in its unbroken or broken phase) through a set of higher-dimensional operators with respect to a mass scale parameter, where the latter represents the energy scale at which NP interactions are generated by an ultra-violet (UV) complete theory, hence allowing one to interpret NP effects in terms of deviations from the SM interactions. Moreover, this is the most powerful method to interpret the impact of correlation effects among operators at different energy scales on cLFV observables.

In the next sections, the main recent literature on these research lines will be reviewed and potential future theoretical prospects will be presented. Wherever possible, the broader picture that connects cLFV with neutrino physics and UV completions of the SM will be discussed. Finally, literature on other relevant cLFV processes, such as the incoherent $\mu\to e$ conversion in nuclei and the neutrinoless double beta decay will also be mentioned.

\section{Precise calculations for experimental backgrounds of cLFV transitions}

The fundamental backgrounds for low-energy cLFV transitions are the radiative and rare leptonic decays, \emph{i.e.} the processes $l\to l'\gamma+2\nu$ and $l\to 3l'+2\nu$, respectively. Indeed, when the energy of the neutrino pair in the final state becomes very small, the two processes are irreducible backgrounds of the forbidden $l\to l'\gamma$ and $l\to 3l'$ decay modes. Hence, the determination of these decay rates is important for the precise determination of the limits on the cLFV BRs.

%\subsection{Radiative leptonic decay}
%
In the last six decades, the radiative decay of the muon has been widely investigated~\cite{Crittenden:1959hm,Eichenberger:1984gi,Pocanic:2014mya,Adam:2013gfn,Arbuzov:2016ywn}; however, the current experimental knowledge is rather unsatisfactory: the Particle Data Group (PDG) value BR$(\mu\to e2\nu\gamma)=(1.4\pm 0.4)\times 10^{-2}$\footnote{Hereafter, to make the quantity well defined, a standard experimental cut on the photon energy $\omega_0=10$ MeV is understood.} is affected by an uncertainty of about $30\%$~\cite{Olive:2016xmw}. The tau lepton and its radiative decays $\tau\to \mu 2\nu\gamma$ and $\tau\to e 2\nu\gamma$ has also been studied by CLEO~\cite{Bergfeld:1999yh}, BaBar~\cite{Lees:2015gea} and Belle~\cite{Abdesselam:2016fmg}, and the current PDG values are BR$(\tau\to\mu 2\nu\gamma)=(3.68\pm 0.10)\times10^{-3}$ and BR$(\tau\to e 2\nu\gamma)=(1.84\pm 0.05)\times 10^{-2}$.

From the theoretical point of view, calculations for the radiative lepton decay are performed in the Fermi theory because full SM corrections are smaller than the NLO QED corrections, hence having no impact on the phenomenological interpretation of current experimental data~\cite{Ferroglia:2013dga,Fael:2013pja}. While tree-level calculations have been available since the 50s of the last century~\cite{Behrends:1955mb,Fronsdal:1959zzb,ECKSTEIN1959297}, it is only recently that full NLO QED computations have been performed, first in~\cite{Fael:2015gua,Fael:2016hnz} and then in~\cite{Pruna:2017upz}.

In particular, in~\cite{Pruna:2017upz} a fully differential Monte Carlo study was presented for the radiative decay of an arbitrarily polarised lepton and maintaining a non-vanishing mass for the lepton of the final state, thus enabling the implementation of arbitrary cuts to mimic closely the experimental framework. The NLO QED matrix elements were computed both in conventional dimensional regularisation\footnote{Exploiting the tools {\tt FeynRules}~\cite{Alloul:2013bka}, {\tt FeynArts}~\cite{Hahn:2000kx}, {\tt FormCalc}~\cite{Hahn:1998yk,Hahn:2016ebn}, {\tt Form}~\cite{Kuipers:2012rf} and {\tt LoopTools}~\cite{Hahn:1998yk}.} and the four-dimensional helicity scheme\footnote{Exploiting a modified version of the program {\tt GoSam}~\cite{Cullen:2014yla}, plus the tools {\tt Ninja}~\cite{Mastrolia:2012bu,Peraro:2014cba,vanDeurzen:2013saa}, {\tt golem95}~\cite{Binoth:2008uq,Mastrolia:2010nb} and {\tt OneLOop}~\cite{vanHameren:2009dr,vanHameren:2010cp}.} (FDH)~\cite{Bern:1991aq}, and full agreement was found for the final physical result\footnote{The equivalence between the two schemes is described in~\cite{Signer:2008va}. See also~\cite{Gnendiger:2017pys} and references therein for a recent review on regularisation schemes.}. On top of this, the on-shell scheme was adopted to renormalise masses and couplings, and soft infra-red (IR) singularities were treated by means of the Frixione--Kunszt--Signer (FKS) subtraction method~\cite{Frixione:1995ms,Frederix:2009yq}.

Apart from the precise calculation of background for tailored studies at specific experiments, other phenomenological issues are clarified by this computation: in~\cite{Fael:2015gua}, a discrepancy of $\sim 3.5\sigma$ was found between the theoretical prediction and the BaBar result for the tau radiative decay in association with an electron. The crucial point is that this measurement was done using rather stringent cuts on the final state and then converted to the standard cut. However, the sizeable terms proportional to $\log{\left(m_e/m_\tau\right)}$ of the QED NLO correction play a fundamental role in such conversion, and~\cite{Pruna:2017upz} shows that their systematic inclusion may reduce the tension to $\sim 1\sigma$.

%\subsection{Rare leptonic decay}
%
The only experimental information available for the rare leptonic decay regards the muonic $\mu\to 3e+2\nu$ transition measured long ago by the SINDRUM collaboration~\cite{Bertl:1985mw} and the $\tau\to 3e+2\nu$ measured by CLEO~\cite{Alam:1995mt}, and their corresponding values according to the PDG~\cite{Olive:2016xmw} are BR$(\mu\to 3e+2\nu)=(3.4\pm 0.4)\times 10^{-5}$ and BR$(\tau\to 3e+\nu)=(2.8\pm 1.5)\times 10^{-5}$.

As regards the theoretical approach to the estimation of the BRs, in complete analogy with the radiative decay, the calculation is performed in the framework of the Fermi theory. On top of tree-level studies~\cite{Fishbane:1985xz,Kersch:1987dw,Dicus:1994dt,Flores-Tlalpa:2015vga}, recently the QED NLO corrections also became available~\cite{Fael:2016yle,Pruna:2016spf}.

Specifically, in~\cite{Pruna:2016spf} the rare decay was computed for a polarised muon in the initial state and maintaining a non-vanishing mass for the lepton of the final state by means of a fully differential Monte Carlo program. The NLO QED matrix elements were computed in the FDH scheme and soft IR singularities were treated by means of the FKS subtraction method, as described in the previous subsection.

The most important outcome of such computation concerns the differential distribution of the BR with respect to the invisible energy $E_{inv}$ carried by the neutrino pair. This distribution displays an interesting behaviour in the limit $E_{inv}\to 0$, which is the region where the background mimics the cLFV signals; here, the NLO QED corrections are negative and approaching $\sim -10\%$. Hence, there are substantially fewer background events than expected from the simplistic tree-level computation, and this has a favourable impact on the efficiency of the searches for cLFV signals.

\section{EFTs for cLFV observables}

In the absence of any evidence for NP at high-energy colliders, the EFT approach to BSM physics is becoming more and more popular in the particle physics community. The SMEFT or analogous setups with the inclusion of higher-dimensional operators~\cite{Petcov:1976ff,Minkowski:1977sc,Weinberg:1979sa,Buchmuller:1985jz,Grzadkowski:2010es,Lehman:2014jma} has become a standard mainstream approach adopted for phenomenological studies of the Higgs boson~\cite{Contino:2013kra,deFlorian:2016spz,Passarino:2016pzb}, neutrino sector~\cite{Elgaard-Clausen:2017xkq}, $B$-physics~\cite{Feruglio:2016gvd,Aebischer:2017gaw,Bordone:2017anc}, \emph{et cetera}.

In the past few years, a lot of effort has also been devoted to including the study of cLFV observables in a consistent EFT approach with higher-dimensional operators. In this connection, the one-loop matching of a subset of dimension-six operators with the $l\to l'\gamma$ dipole interaction was performed in~\cite{Crivellin:2013hpa}, then the complete set was considered in~\cite{Pruna:2014asa} where also renormalisation-group evolution (RGE) effects of the operators were taken into account in a systematic way\footnote{An independent calculation of the complete set of SMEFT operators anomalous dimensions was presented in~\cite{Jenkins:2013zja,Jenkins:2013wua,Alonso:2013hga}.}. Consequently, it was possible to interpret the MEG limits on BR$(\mu\to e\gamma)$~\cite{Adam:2013mnn} and the BaBar/Belle limit on BR$(\tau\to e(\mu)\gamma)$~\cite{Hayasaka:2007vc,Aubert:2009ag} in terms of new constraints on the parameter space of some SMEFT Wilson coefficients defined at the high-energy scale that were previously unbounded by low-energy experiments.

Thereafter, the impact of specific low- and high-energy searches was compared in~\cite{Pruna:2015jhf}. In this connection, it was found that LFV decays of the $Z$-boson (\emph{i.e.} $Z\to l_1 l_2$) are much more constrained from low-energy experiments than from the limits of current (and possibly future, as discussed also in~\cite{Abada:2014cca}) direct searches at high energy.

Then, the study of cLFV within the framework of a QED EFT\footnote{Exploiting the same operatorial basis introduced by~\cite{Kuno:1999jp,Cirigliano:2009bz}.} was performed systematically in a one-loop\footnote{In addition, specific leading order two-loop terms of the RGE were included to provide important qualitative new effects. Previous constraints on Wilson coefficients have been extracted at the scale of the experiments for cLFV lepton--quark~\cite{Davidson:1993qk,Carpentier:2010ue}, lepton--tau~\cite{Dassinger:2007ru,Celis:2014asa}, four-lepton~\cite{Raidal:2008jk} and lepton--gluon~\cite{Petrov:2013vka} operators. Furthermore, in~\cite{Cirigliano:2017azj}, the impact of LFV tensor and axial-vector four-fermion operators that couple to the spin of nucleons was studied.} RGE-improved scenario~\cite{Davidson:2016edt,Crivellin:2016ebg,Crivellin:2017rmk}. Focusing on the muonic sector, the three processes $\mu\to e\gamma$, $\mu\to 3e$ and coherent nuclear $\mu\to e$ conversion were studied considering the RGE of the Wilson coefficients between the electroweak and the experimental scale ($\Lambda=m_\mu$ for the muonic decays and $\Lambda\simeq 1$ GeV for the conversion in nuclei). As a result, muonic decay and conversion rates were interpreted as functions of the Wilson coefficients at any scale up to $m_W$. Taking the experimental limits on these processes as input, it was found that a considerable set of Wilson coefficients unbounded in the tree-level approach were instead severely constrained. In addition, correlations among operators were studied both in the light of current data and future experimental prospects, illustrating the complementarity of searches planned for the MEG~II~\cite{Baldini:2013ke}, Mu3e~\cite{Blondel:2013ia} and COMET/Mu2e~\cite{Cui:2009zz,Abrams:2012er} experiments.

Keeping in mind the ultimate goal of the EFT approach, it is worth mentioning that a plethora of UV-complete models designed for the most different reasons can be matched effectively to cLFV operators and become part of the discussed analysis\footnote{See~\cite{Pascoli:2016wlt,Geib:2015tvt,Geib:2015unm} for model setups that are favourable to the interpretation of neutrino observables.}.

Finally, one should also bear in mind that the dimension-up-to-six operatorial setup is not sufficient to include lepton-number violation in the systematic approach described above, and more should be done to effectively describe UV-complete scenarios that provide incoherent $\mu\to e$ conversion in nuclei~\cite{Domin:2004tk,Geib:2016atx,Geib:2016daa} and the neutrinoless double beta decay~\cite{Abada:2014nwa,Deppisch:2014zta,Deppisch:2017vne}.

\section{Conclusion}

This document reviewed recent developments and literature concerning cLFV in the context of precise calculations for irreducible backgrounds and EFT interpretation of the current and future outcomes of cLFV experiments.

\Acknowledgements
\noindent
I am grateful to Adrian Signer and Yannick Ulrich for useful comments while writing this manuscript.

\newpage
\bibliographystyle{woc} 
{\small   
\bibliography{nuphys}

\begin{thebibliography}{102}

\bibitem{TheMEG:2016wtm}
A.M. Baldini et~al. (MEG), Eur. Phys. J. \textbf{C76}, 434 (2016),
  \texttt{1605.05081}

\bibitem{Bellgardt:1987du}
U.~Bellgardt et~al. (SINDRUM), Nucl. Phys. \textbf{B299}, 1 (1988)

\bibitem{Bolton:1988af}
R.D. Bolton et~al., Phys. Rev. \textbf{D38}, 2077 (1988)

\bibitem{Bertl:2006up}
W.H. Bertl et~al. (SINDRUM II), Eur. Phys. J. \textbf{C47}, 337 (2006)

\bibitem{Aubert:2009ag}
B.~Aubert et~al. (BaBar), Phys. Rev. Lett. \textbf{104}, 021802 (2010),
  \texttt{0908.2381}

\bibitem{Hayasaka:2007vc}
K.~Hayasaka et~al. (Belle), Phys. Lett. \textbf{B666}, 16 (2008),
  \texttt{0705.0650}

\bibitem{Lees:2010ez}
J.P. Lees et~al. (BaBar), Phys. Rev. \textbf{D81}, 111101 (2010),
  \texttt{1002.4550}

\bibitem{Aaij:2014azz}
R.~Aaij et~al. (LHCb), JHEP \textbf{02}, 121 (2015), \texttt{1409.8548}

\bibitem{Aad:2014bca}
G.~Aad et~al. (ATLAS), Phys. Rev. \textbf{D90}, 072010 (2014),
  \texttt{1408.5774}

\bibitem{Akers:1995gz}
R.~Akers et~al. (OPAL), Z. Phys. \textbf{C67}, 555 (1995)

\bibitem{Abreu:1996mj}
P.~Abreu et~al. (DELPHI), Z. Phys. \textbf{C73}, 243 (1997)

\bibitem{Khachatryan:2015kon}
V.~Khachatryan et~al. (CMS), Phys. Lett. \textbf{B749}, 337 (2015),
  \texttt{1502.07400}

\bibitem{Fael:2015gua}
M.~Fael, L.~Mercolli, M.~Passera, JHEP \textbf{07}, 153 (2015),
  \texttt{1506.03416}

\bibitem{Fael:2016hnz}
M.~Fael, M.~Passera (2016), \texttt{1602.00457}

\bibitem{Pruna:2017upz}
G.M. Pruna, A.~Signer, Y.~Ulrich (2017), \texttt{1705.03782}

\bibitem{Fael:2016yle}
M.~Fael, C.~Greub, JHEP \textbf{01}, 084 (2017), \texttt{1611.03726}

\bibitem{Pruna:2016spf}
G.M. Pruna, A.~Signer, Y.~Ulrich, Phys. Lett. \textbf{B765}, 280 (2017),
  \texttt{1611.03617}

\bibitem{Baldini:2013ke}
A.M. Baldini et~al. (2013), \texttt{1301.7225}

\bibitem{Blondel:2013ia}
A.~Blondel et~al. (2013), \texttt{1301.6113}

\bibitem{Abe:2010gxa}
T.~Abe et~al. (Belle-II) (2010), \texttt{1011.0352}

\bibitem{Pruna:2014asa}
G.M. Pruna, A.~Signer, JHEP \textbf{10}, 014 (2014), \texttt{1408.3565}

\bibitem{Pruna:2015jhf}
G.M. Pruna, A.~Signer, EPJ Web Conf. \textbf{118}, 01031 (2016),
  \texttt{1511.04421}

\bibitem{Crivellin:2016ebg}
A.~Crivellin, S.~Davidson, G.M. Pruna, A.~Signer (2016), \texttt{1611.03409}

\bibitem{Crivellin:2017rmk}
A.~Crivellin, S.~Davidson, G.M. Pruna, A.~Signer (2017), \texttt{1702.03020}

\bibitem{Crittenden:1959hm}
R.R. Crittenden, W.D. Walker, J.~Ballam, Phys. Rev. \textbf{121}, 1823 (1961)

\bibitem{Eichenberger:1984gi}
W.~Eichenberger, R.~Engfer, A.~Van Der~Schaaf, Nucl. Phys. \textbf{A412}, 523
  (1984)

\bibitem{Pocanic:2014mya}
D.~Pocanic et~al., Int. J. Mod. Phys. Conf. Ser. \textbf{35}, 1460437 (2014),
  \texttt{1403.7416}

\bibitem{Adam:2013gfn}
A.M. Baldini et~al. (MEG), Eur. Phys. J. \textbf{C76}, 108 (2016),
  \texttt{1312.3217}

\bibitem{Arbuzov:2016ywn}
A.B. Arbuzov, T.V. Kopylova, JHEP \textbf{09}, 109 (2016), \texttt{1605.06612}

\bibitem{Olive:2016xmw}
C.~Patrignani et~al. (Particle Data Group), Chin. Phys. \textbf{C40}, 100001
  (2016)

\bibitem{Bergfeld:1999yh}
T.~Bergfeld et~al. (CLEO), Phys. Rev. Lett. \textbf{84}, 830 (2000),
  \texttt{hep-ex/9909050}

\bibitem{Lees:2015gea}
J.P. Lees et~al. (BaBar), Phys. Rev. \textbf{D91}, 051103 (2015),
  \texttt{1502.01784}

\bibitem{Abdesselam:2016fmg}
A.~Abdesselam et~al. (Belle) (2016), \texttt{1609.08280}

\bibitem{Ferroglia:2013dga}
A.~Ferroglia, C.~Greub, A.~Sirlin, Z.~Zhang, Phys. Rev. \textbf{D88}, 033012
  (2013), \texttt{1307.6900}

\bibitem{Fael:2013pja}
M.~Fael, L.~Mercolli, M.~Passera, Phys. Rev. \textbf{D88}, 093011 (2013),
  \texttt{1310.1081}

\bibitem{Behrends:1955mb}
R.E. Behrends, R.J. Finkelstein, A.~Sirlin, Phys. Rev. \textbf{101}, 866 (1956)

\bibitem{Fronsdal:1959zzb}
C.~Fronsdal, H.~Uberall, Phys. Rev. \textbf{113}, 654 (1959)

\bibitem{ECKSTEIN1959297}
S.~Eckstein, R.~Pratt, Annals of Physics \textbf{8}, 297  (1959)

\bibitem{Alloul:2013bka}
A.~Alloul, N.D. Christensen, C.~Degrande, C.~Duhr, B.~Fuks, Comput. Phys.
  Commun. \textbf{185}, 2250 (2014), \texttt{1310.1921}

\bibitem{Hahn:2000kx}
T.~Hahn, Comput. Phys. Commun. \textbf{140}, 418 (2001),
  \texttt{hep-ph/0012260}

\bibitem{Hahn:1998yk}
T.~Hahn, M.~Perez-Victoria, Comput. Phys. Commun. \textbf{118}, 153 (1999),
  \texttt{hep-ph/9807565}

\bibitem{Hahn:2016ebn}
T.~Hahn, S.~Paßehr, C.~Schappacher, PoS \textbf{LL2016}, 068 (2016), [J. Phys.
  Conf. Ser.762,no.1,012065(2016)], \texttt{1604.04611}

\bibitem{Kuipers:2012rf}
J.~Kuipers, T.~Ueda, J.A.M. Vermaseren, J.~Vollinga, Comput. Phys. Commun.
  \textbf{184}, 1453 (2013), \texttt{1203.6543}

\bibitem{Cullen:2014yla}
G.~Cullen et~al., Eur. Phys. J. \textbf{C74}, 3001 (2014), \texttt{1404.7096}

\bibitem{Mastrolia:2012bu}
P.~Mastrolia, E.~Mirabella, T.~Peraro, JHEP \textbf{06}, 095 (2012), [Erratum:
  JHEP11,128(2012)], \texttt{1203.0291}

\bibitem{Peraro:2014cba}
T.~Peraro, Comput. Phys. Commun. \textbf{185}, 2771 (2014), \texttt{1403.1229}

\bibitem{vanDeurzen:2013saa}
H.~van Deurzen, G.~Luisoni, P.~Mastrolia, E.~Mirabella, G.~Ossola, T.~Peraro,
  JHEP \textbf{03}, 115 (2014), \texttt{1312.6678}

\bibitem{Binoth:2008uq}
T.~Binoth, J.P. Guillet, G.~Heinrich, E.~Pilon, T.~Reiter, Comput. Phys.
  Commun. \textbf{180}, 2317 (2009), \texttt{0810.0992}

\bibitem{Mastrolia:2010nb}
P.~Mastrolia, G.~Ossola, T.~Reiter, F.~Tramontano, JHEP \textbf{08}, 080
  (2010), \texttt{1006.0710}

\bibitem{vanHameren:2009dr}
A.~van Hameren, C.G. Papadopoulos, R.~Pittau, JHEP \textbf{09}, 106 (2009),
  \texttt{0903.4665}

\bibitem{vanHameren:2010cp}
A.~van Hameren, Comput. Phys. Commun. \textbf{182}, 2427 (2011),
  \texttt{1007.4716}

\bibitem{Bern:1991aq}
Z.~Bern, D.A. Kosower, Nucl. Phys. \textbf{B379}, 451 (1992)

\bibitem{Signer:2008va}
A.~Signer, D.~Stockinger, Nucl. Phys. \textbf{B808}, 88 (2009),
  \texttt{0807.4424}

\bibitem{Gnendiger:2017pys}
C.~Gnendiger et~al. (2017), \texttt{1705.01827}

\bibitem{Frixione:1995ms}
S.~Frixione, Z.~Kunszt, A.~Signer, Nucl. Phys. \textbf{B467}, 399 (1996),
  \texttt{hep-ph/9512328}

\bibitem{Frederix:2009yq}
R.~Frederix, S.~Frixione, F.~Maltoni, T.~Stelzer, JHEP \textbf{10}, 003 (2009),
  \texttt{0908.4272}

\bibitem{Bertl:1985mw}
W.H. Bertl et~al. (SINDRUM), Nucl. Phys. \textbf{B260}, 1 (1985)

\bibitem{Alam:1995mt}
M.S. Alam et~al. (CLEO), Phys. Rev. Lett. \textbf{76}, 2637 (1996)

\bibitem{Fishbane:1985xz}
P.M. Fishbane, K.J.F. Gaemers, Phys. Rev. \textbf{D33}, 159 (1986)

\bibitem{Kersch:1987dw}
A.~Kersch, N.~Kraus, R.~Engfer, Nucl. Phys. \textbf{A485}, 606 (1988)

\bibitem{Dicus:1994dt}
D.A. Dicus, R.~Vega, Phys. Lett. \textbf{B338}, 341 (1994),
  \texttt{hep-ph/9402262}

\bibitem{Flores-Tlalpa:2015vga}
A.~Flores-Tlalpa, G.~López~Castro, P.~Roig, JHEP \textbf{04}, 185 (2016),
  \texttt{1508.01822}

\bibitem{Petcov:1976ff}
S.T. Petcov, Sov. J. Nucl. Phys. \textbf{25}, 340 (1977), [Erratum: Yad.
  Fiz.25,1336(1977)]

\bibitem{Minkowski:1977sc}
P.~Minkowski, Phys. Lett. \textbf{B67}, 421 (1977)

\bibitem{Weinberg:1979sa}
S.~Weinberg, Phys. Rev. Lett. \textbf{43}, 1566 (1979)

\bibitem{Buchmuller:1985jz}
W.~Buchmuller, D.~Wyler, Nucl. Phys. \textbf{B268}, 621 (1986)

\bibitem{Grzadkowski:2010es}
B.~Grzadkowski, M.~Iskrzynski, M.~Misiak, J.~Rosiek, JHEP \textbf{10}, 085
  (2010), \texttt{1008.4884}

\bibitem{Lehman:2014jma}
L.~Lehman, Phys. Rev. \textbf{D90}, 125023 (2014), \texttt{1410.4193}

\bibitem{Contino:2013kra}
R.~Contino, M.~Ghezzi, C.~Grojean, M.~Muhlleitner, M.~Spira, JHEP \textbf{07},
  035 (2013), \texttt{1303.3876}

\bibitem{deFlorian:2016spz}
D.~de~Florian et~al. (LHC Higgs Cross Section Working Group) (2016),
  \texttt{1610.07922}

\bibitem{Passarino:2016pzb}
G.~Passarino, M.~Trott (2016), \texttt{1610.08356}

\bibitem{Elgaard-Clausen:2017xkq}
G.~Elgaard-Clausen, M.~Trott (2017), \texttt{1703.04415}

\bibitem{Feruglio:2016gvd}
F.~Feruglio, P.~Paradisi, A.~Pattori, Phys. Rev. Lett. \textbf{118}, 011801
  (2017), \texttt{1606.00524}

\bibitem{Aebischer:2017gaw}
J.~Aebischer, M.~Fael, C.~Greub, J.~Virto (2017), \texttt{1704.06639}

\bibitem{Bordone:2017anc}
M.~Bordone, G.~Isidori, S.~Trifinopoulos (2017), \texttt{1702.07238}

\bibitem{Crivellin:2013hpa}
A.~Crivellin, S.~Najjari, J.~Rosiek, JHEP \textbf{04}, 167 (2014),
  \texttt{1312.0634}

\bibitem{Jenkins:2013zja}
E.E. Jenkins, A.V. Manohar, M.~Trott, JHEP \textbf{10}, 087 (2013),
  \texttt{1308.2627}

\bibitem{Jenkins:2013wua}
E.E. Jenkins, A.V. Manohar, M.~Trott, JHEP \textbf{01}, 035 (2014),
  \texttt{1310.4838}

\bibitem{Alonso:2013hga}
R.~Alonso, E.E. Jenkins, A.V. Manohar, M.~Trott, JHEP \textbf{04}, 159 (2014),
  \texttt{1312.2014}

\bibitem{Adam:2013mnn}
J.~Adam et~al. (MEG), Phys. Rev. Lett. \textbf{110}, 201801 (2013),
  \texttt{1303.0754}

\bibitem{Abada:2014cca}
A.~Abada, V.~De~Romeri, S.~Monteil, J.~Orloff, A.M. Teixeira, JHEP \textbf{04},
  051 (2015), \texttt{1412.6322}

\bibitem{Kuno:1999jp}
Y.~Kuno, Y.~Okada, Rev. Mod. Phys. \textbf{73}, 151 (2001),
  \texttt{hep-ph/9909265}

\bibitem{Cirigliano:2009bz}
V.~Cirigliano, R.~Kitano, Y.~Okada, P.~Tuzon, Phys. Rev. \textbf{D80}, 013002
  (2009), \texttt{0904.0957}

\bibitem{Davidson:1993qk}
S.~Davidson, D.C. Bailey, B.A. Campbell, Z. Phys. \textbf{C61}, 613 (1994),
  \texttt{hep-ph/9309310}

\bibitem{Carpentier:2010ue}
M.~Carpentier, S.~Davidson, Eur. Phys. J. \textbf{C70}, 1071 (2010),
  \texttt{1008.0280}

\bibitem{Dassinger:2007ru}
B.M. Dassinger, T.~Feldmann, T.~Mannel, S.~Turczyk, JHEP \textbf{10}, 039
  (2007), \texttt{0707.0988}

\bibitem{Celis:2014asa}
A.~Celis, V.~Cirigliano, E.~Passemar, Phys. Rev. \textbf{D89}, 095014 (2014),
  \texttt{1403.5781}

\bibitem{Raidal:2008jk}
M.~Raidal et~al., Eur. Phys. J. \textbf{C57}, 13 (2008), \texttt{0801.1826}

\bibitem{Petrov:2013vka}
A.A. Petrov, D.V. Zhuridov, Phys. Rev. \textbf{D89}, 033005 (2014),
  \texttt{1308.6561}

\bibitem{Cirigliano:2017azj}
V.~Cirigliano, S.~Davidson, Y.~Kuno (2017), \texttt{1703.02057}

\bibitem{Davidson:2016edt}
S.~Davidson, Eur. Phys. J. \textbf{C76}, 370 (2016), \texttt{1601.07166}

\bibitem{Cui:2009zz}
Y.G. Cui et~al. (COMET) (2009), KEK-2009-10

\bibitem{Abrams:2012er}
R.J. Abrams et~al. (Mu2e) (2012), \texttt{1211.7019}

\bibitem{Pascoli:2016wlt}
S.~Pascoli, Y.L. Zhou, JHEP \textbf{10}, 145 (2016), \texttt{1607.05599}

\bibitem{Geib:2015tvt}
T.~Geib, S.F. King, A.~Merle, J.M. No, L.~Panizzi, Phys. Rev. \textbf{D93},
  073007 (2016), \texttt{1512.04391}

\bibitem{Geib:2015unm}
T.~Geib, A.~Merle, Phys. Rev. \textbf{D93}, 055039 (2016), \texttt{1512.04225}

\bibitem{Domin:2004tk}
P.~Domin, S.~Kovalenko, A.~Faessler, F.~Simkovic, Phys. Rev. \textbf{C70},
  065501 (2004), \texttt{nucl-th/0409033}

\bibitem{Geib:2016atx}
T.~Geib, A.~Merle, K.~Zuber, Phys. Lett. \textbf{B764}, 157 (2017),
  \texttt{1609.09088}

\bibitem{Geib:2016daa}
T.~Geib, A.~Merle, Phys. Rev. \textbf{D95}, 055009 (2017), \texttt{1612.00452}

\bibitem{Abada:2014nwa}
A.~Abada, V.~De~Romeri, A.M. Teixeira, JHEP \textbf{09}, 074 (2014),
  \texttt{1406.6978}

\bibitem{Deppisch:2014zta}
F.F. Deppisch, T.E. Gonzalo, S.~Patra, N.~Sahu, U.~Sarkar, Phys. Rev.
  \textbf{D91}, 015018 (2015), \texttt{1410.6427}

\bibitem{Deppisch:2017vne}
F.F. Deppisch, C.~Hati, S.~Patra, P.~Pritimita, U.~Sarkar (2017),
  \texttt{1701.02107}

\end{thebibliography}
}
 
\end{document}